\documentclass[12pt]{article}
\pdfoutput=1
\usepackage{tikz}
\usetikzlibrary{matrix,arrows,decorations.pathmorphing,decorations.pathreplacing}
\usepackage{jheppub}
\usepackage{amsmath,amssymb,euscript,array,mathrsfs,appendix,ctable,marvosym}
\usepackage{arydshln}
\usepackage{todonotes}
\usepackage{graphicx}
\usepackage[normalem]{ulem}

\input{tcilatex}

\newcommand{\EQ}[1]{\begin{equation}\begin{split} #1
\end{split}\end{equation}}

\title{$\lambda$-Deformation Of The $AdS_{5}\times S^{5}$ Pure Spinor Superstring}
\author[a]{H\'ector A. Ben\'itez,\footnote{e-mail: hectorbenitez@usp.br}}
\author[b]{and David M. Schmidtt.\footnote{e-mail: david@df.ufscar.br}}

\affiliation[a]{Instituto de F\'\i sica, Universidade de S\~ao Paulo, \\
Rua do Mat\~ao, 1371, 05508-090, S\~ao Paulo, SP, Brasil}
\affiliation[b]{Departamento de F\'\i sica, Universidade Federal de S\~ao Carlos, \\
Caixa Postal 676, CEP 13565-905, S\~ao Carlos, SP, Brasil} 

\abstract{The lambda deformation of the pure spinor formalism of the superstring in the $AdS_{5}\times S^{5}$ background is introduced. It is shown that the deformation preserves the integrability as well as the one-loop conformal invariance of its parent theory. It is also shown that the effective action takes the standard form of the Berkovits-Howe action functional, allowing to calculate the deformed background supergeometry in a straightforward way. The background fields coincide with those of the lambda model of the Green-Schwarz formalism, hence satisfying the same set of supergravity equations of motion.  \\
\begin{flushleft}
\end{flushleft}
}
\begin{document}

\maketitle


\section{Introduction}

A fundamental characteristic of the gravity sector of the AdS/CFT correspondence is that superstrings propagating in the $AdS_{5}\times S^{5}$ background can be described by a (classsical) integrable model \cite{BPR}. This
geometry rises as a solution of the type IIB supergravity equations of motion when supported by a self-dual
Ramond-Ramond (RR) five-form flux. It is well known that a superstring moving on a curved background including a RR
flux can be correctly formulated by either the Green-Schwarz (GS) formalism \cite{MT} or the pure spinor (PS)
formalism \cite{BC} and that both formulations present manifest target-space supersymmetry. In both formulations, classical integrability is ensured since
the equations of motion can be cast into a zero curvature equation satisfied by a Lax connection, see \cite{BPR,breno}. Although the complete quantization of
the superstring in the $AdS_{5}\times S^{5}$ geometry has never been fulfilled and still remains as an open problem, the use of integrability techniques yield significant progress in understanding the excitation spectrum (see \cite{Arutyunov:2009ga} for a review). It is then reasonable to believe that integrability could play a prominent role on an eventual first principle quantization approach of the theory.

Concerning the GS formalism, in the last years, two different but complementary types of integrable deformations of the GS $AdS_{5}\times S^{5}$ superstring, have attracted a considerable deal of attention. On one hand we have the Yang-Baxter (YB) deformations, so named because they are characterized by a linear operator acting on the Lie superalgebra $\mathfrak{psu}(2,2|4)$, which solves the modified classical Yang-Baxter equation mCYBE. This deformation was first introduced for the principal chiral model by Klim\v{c}\'ik in \cite{Klimcik} and subsequenty developed in \cite{eta bos,eta-fer1,eta-fer2} for (super)-strings on (semi)-symmetric spaces (see \cite{kawa,matsu} as well). In order to provide a target space which solves the equations of motion of type IIB supergravity, these R-matrices must satisfy the unimodular condition \cite{Borsato}. On the other hand we have the lambda deformations, which are based on a $G/G$ gauged WZW model and are better understood as deformations of the non-Abelian T-dual of the original theory, they were first introduced for the principal chiral model by Sfetsos in \cite{Sfetsos} and subsequently developed in \cite{lambda-bos,quantum-group,lambda-fer,lambda background} for (super)-strings on (semi)-symmetric spaces. An important fact of the lambda deformation is that it produces string theory backgrounds solving the equations of motion of type IIB supergravity, see \cite{lambda back,lambda ads3xs3, Borsato,Sfetsos2}. Although each family of deformations produce different target space supergeometries, it has been proposed that, through an analytic continuation, the lambda deformations are Poisson-Lie T-dual to the YB deformations with R-matrices satisfying the non-split mCYBE \cite{K,V}. In relation to the classical stringy configurations on $\eta$ and $\lambda$ backgrounds, integrability conditions are discussed in \cite{Roychowdhury}.
  
In respect to the PS formalism, integrable deformations of the $AdS_{5}\times S^{5}$ superstring have received much less attention and this is an unsatisfactory scenario for many reasons. For instance, in the PS formalism the world-sheet metric is already in the conformal gauge and
the problematic $\kappa$-symmetry, signature of the GS superstring, is replaced by a global and much better behaved BRST symmetry,
therefore avoiding delicate issues involving the light-cone gauge not to mention the lack of a satisfactory covariant quantization scheme due to the fact that the first and second class fermionic constraints cannot be disentangled covariantly in the GS formalism (a problem not present in the PS formalism by construction). Recently, the homogeneous YB deformations of the PS superstring were introduced in \cite{eta-PS}, by following the homological perturbation theory developed in \cite{Bevilaqua}. It was shown that its target space background turns out to be the same found for the YB deformation of the GS superstring \cite{Borsato}. From the PS point of view, the deformed space solving the type IIB supergravity equations of motion is produced by a particular set of primary vertex operators belonging to the BRST cohomology in $AdS_{5}\times S^{5}$. In this context, the mCYBE condition on the $R$-matrices, needed for integrability of the deformed action, arises by imposing the nilpotency of the deformed BRST charge, revealing a profound connection between the integrability of the deformed theory and its BRST symmetry.

With this vision, in this paper we introduce the lambda deformation of the $AdS_{5}\times S^{5}$ PS superstring. The deformation preserves the main characteristics of the un-deformed theory: its BRST symmetry, its classical integrability, their local symmetries and its conformal symmetry at one-loop. In addition, it describes exactly the same supergeometry associated to its lambda deformed GS counterpart, much in the same way as the YB deformations of the GS and PS formulations describe the same background and are equivalent as string theories, at least, from the classical theory point of view.

The paper is organized as follows. In section \eqref{2}, we introduce the action functional of the lambda deformed $AdS_{5}\times S^{5}$ pure spinor superstring and consider the equations of motion, the classical integrability and the BRST symmetry of the theory. In section \eqref{3}, we run the Dirac procedure and study the integrability of the deformed theory from the Hamiltonian theory point of view. The analysis is simpler than in the GS formalism because of the absence of kappa symmetry and is essentially the same of the lambda deformed hybrid superstring. It is shown that the classical exchange algebra for the spatial component of the Lax connection takes Maillet $\mathfrak{r}/\mathfrak{s}$ form, as expected. In section \eqref{4}, we consider the quantum conformal symmetry and compute the one-loop beta function of the deformed theory and show that it vanishes. Finally, we show that the effective action for the deformed theory can be cast into the standard form of the Berkovits-Howe action functional. Once in this form, the deformed target space fields can be easily identified. They are exactly the same as the ones entering the geometry of the lambda deformed GS superstring, meaning that both formulations describe the same classical system. 

\section{Lambda deformed pure spinor superstring}\label{2}

In this section we recall the lambda deformed action of the PS superstring in the $AdS_{5}\times S^{5}$ background originally constructed in \cite{exploring}. It is shown that the deformation preserves the integrability of its parent theory, a discussion not covered previously in \cite{exploring}. The main characteristic of the deformed theory integrable structure is that its associated Lax connection depends explicitly on the deformation parameter $\lambda$, a feature not observed before in any of the known lambda models. BRST symmetry is briefly considered as well from the symplectic theory point of view.

\subsection{Action functional}

Consider the Lie superalgebra $\mathfrak{f}=\mathfrak{psu}(2,2|4)$ and its $\mathbb{Z}_{4}$ decomposition induced by the automorphism $\Phi $%
\begin{equation}
\Phi (\mathfrak{f}^{(j)})={\textbf{i}}^{j}\mathfrak{f}^{(j)},\text{ \ \ }\mathfrak{f=}%
\bigoplus\nolimits_{i=0}^{3}\mathfrak{f}^{(i)},\text{ \ \ }[\mathfrak{f}%
^{(i)},\mathfrak{f}^{(j)}]\subset \mathfrak{f}^{(i+j)\func{mod}4}.\text{ \ \ 
}  \label{auto}
\end{equation}%
From this decomposition we associate a twisted loop superagebra given by
\begin{equation}
\hat{\mathfrak{f}}=\bigoplus\nolimits_{n\in 
\mathbb{Z}
}\left( \bigoplus\nolimits_{i=0}^{3}\mathfrak{f}^{(i)}\otimes
z^{4n+i}\right) =\bigoplus\nolimits_{n\in 
\mathbb{Z}
}\hat{\mathfrak{f}}^{(n)},  \label{loop superalgebra}
\end{equation}
required later on for describing the integrable structure of the field theory, where $z$ plays the role of the spectral parameter.

The lambda deformation of the $AdS_{5}\times S^{5}$ pure spinor (PS) superstring is defined by the following action functional\footnote{%
The 1+1 notation used is: $\sigma^{\pm }=\tau\pm \sigma,$ $\partial
_{\pm }=\frac{1}{2}(\partial _{\tau}\pm \partial _{\sigma}),$ $\eta _{\mu \nu
}=diag(1,-1)$, $\epsilon ^{01}=1$, $\delta _{\sigma\sigma ^{\prime }}$=$\delta(\sigma-\sigma^{\prime})$ and $\delta^{\prime} _{\sigma\sigma ^{\prime }}$=$\partial_{\sigma}\delta(\sigma-\sigma^{\prime})$. Also $a_{\pm }=\frac{1}{2}%
(a_{\tau}\pm a_{\sigma})$ and sometimes we use $\tau=\sigma^{0}$ and $\sigma=\sigma^{1}$ interchangeably.}\cite{exploring}
\begin{equation}
S_{\lambda}=S_{\Omega}+S_{PS}. \label{deformed PS}
\end{equation}

The first contribution is given by the matter sector
\begin{equation}
S_{\Omega}=S_{F/F}(\mathcal{F},A_{\pm })-\frac{k}{\pi }\dint_{\Sigma
}d^{2}\sigma \left\langle A_{+}(\Omega -1)A_{-}\right\rangle ,\text{ \ \ }k\in \mathbb{Z} \label{deformed-Hybrid},  
\end{equation}%
where $\left\langle \ast ,\ast \right\rangle= str(\ast ,\ast )$ is the
supertrace in some faithful representation of the Lie superalgebra $\mathfrak{f}$, $\Sigma=\mathbb{R}\times S^{1} $ is the world-sheet manifold parameterized by the coordinates $(\tau, \sigma)$ and $\Omega\equiv \Omega(\lambda)$, where
\begin{equation}
\Omega(z)=P^{(0)}+z^{-3}P^{(1)}+z^{-2}P^{(2)}+z^{-1}P^{(3)}, \text{ \ \ } \lambda^{-2}=1+\frac{\kappa^{2}}{k}
\end{equation}
is the omega projector that defines the lambda deformation of the hybrid superstring \cite{exploring}. Above, we have that%
\begin{equation}
S_{F/F}(\mathcal{F},A_{\mu })=S_{WZW}(\mathcal{F})_{k}-\frac{k}{\pi }%
\dint_{\Sigma }d^{2}\sigma \left\langle A_{+}\partial _{-}\mathcal{FF}^{-1}-A_{-}%
\mathcal{F}^{-1}\partial _{+}\mathcal{F-}A_{+}\mathcal{F}A_{-}\mathcal{F}%
^{-1}+A_{+}A_{-}\right\rangle ,
\end{equation}%
where $S_{WZW}(\mathcal{F})_{k}$ is the level $k$ WZW model action
\begin{equation}
S_{WZW}(\mathcal{F})_{k}=-\frac{k}{2\pi }\int\nolimits_{\Sigma }d^{2}\sigma
\left\langle \mathcal{F}^{-1}\partial _{+}\mathcal{FF}^{-1}\partial _{-}%
\mathcal{F}\right\rangle -\frac{k}{4\pi }\int\nolimits_{B}\chi(\mathcal{F}^{\prime}) ,\text{ \ \ }%
\chi(\mathcal{F}^{\prime}) =\frac{1}{3}\langle (\mathcal{F^{\prime}}^{-1}d\mathcal{F^{\prime})}%
^{3}\rangle .
\end{equation}%
The $\kappa^2$ is the coupling constant of the undeformed theory and the $P^{(i)}$ are projectors along the subspaces $\mathfrak{f}^{(i)}\subset \mathfrak{f}$ induced by the $\mathbb{Z}_{4}$ decomposition. The deformation parameter $\lambda$ takes values in $0 \leq \lambda \leq 1$.

The second contribution is given by the ghost sector
\begin{equation}
S_{PS}=-\frac{k}{\pi }(\lambda^{-4}-1)\dint_{\Sigma
}d^{2}\sigma \langle w^{(3)}_{+}D_{-}^{(0)}l^{(1)}+\hat{w}^{(1)}_{-}D_{+}^{(0)}\hat{l}^{(3)}-N^{(0)}_{+}\hat{N}^{(0)}_{-} \rangle , \label{PS-term} 
\end{equation}%
where
\begin{equation}
N^{(0)}_{+}=-[w^{(3)}_{+},l^{(1)}]_{+},\text{ \ \ }\hat{N}^{(0)}_{-}=-[\hat{w}^{(1)}_{-},\hat{l}^{(3)}]_{+} \label {PS Lorentz}
\end{equation}
are the pure spinor Lorentz currents, $(w^{(3)}_{+},\hat{w}^{(1)}_{-})$ are the conjugate fields of the bosonic pure spinor ghosts $(l^{(1)},\hat{l}^{(3)})$ and $D_{\pm}^{(0)}=\partial(*)+[A_{\pm}^{(0)},*]$ is a covariant derivative with respect to the gauge symmetry associated to the subalgebra $\mathfrak{f}^{(0)}\subset \mathfrak{f}$ . The ghosts satisfy the pure spinor conditions 
\begin{equation}
[l^{(1)},l^{(1)}]_{+}=0,\text{ \ \ }[\hat{l}^{(3)},\hat{l}^{(3)}]_{+}=0, \label{PS constraints}
\end{equation} 
where $[*,*]_{+}$ denotes the anti-commutator.

The relation of the action \eqref{deformed PS} and the action of the un-deformed theory is found by considering the $\lambda \rightarrow 1$ limit, which is defined by expanding the group-like Lagrangian field near the identity $\mathcal{F=}1+\kappa ^{2}\nu /k+...$
while taking $k\rightarrow \infty $ and keeping $\kappa ^{2}$ fixed. We find that%
\begin{equation}
\Omega =1+\frac{\kappa ^{2}}{k}\theta +...,\text{ \ \ }\theta =P^{(2)}+\frac{%
3}{2}P^{(1)}+\frac{1}{2}P^{(3)}. 
\end{equation}%
In this limit, the deformed action reduces to the action of the pure spinor
superstring written in the first order (or non-Abelian Buscher) form%
\begin{equation}
S_{PS}=-\frac{\kappa ^{2}}{\pi }\dint_{\Sigma }d^{2}x\left\langle
A_{+}\theta A_{-}+\nu F_{+-}\right\rangle -\frac{2\kappa ^{2}}{\pi }%
\dint_{\Sigma }d^{2}x\big\langle w^{(3)}_{+}D_{-}^{(0)}l^{(1)}+\hat{w}^{(1)}_{-}D_{+}^{(0)}\hat{l}^{(3)}-N^{(0)}_{+}\hat{N}^{(0)}_{-} \rangle+...,
\end{equation}%
where the ellipsis denote sub-leading terms of order $1/k$ and $F_{+-}$ is the field strength of $A_{\pm}$. After using the
equations of motion for the (now) Lagrange multiplier field $\nu $ and by fixing the
gauge $A_{\pm }=J_{\pm }=f^{-1}\partial _{\pm }f,$ we recover the pure spinor superstring action functional. If we instead integrate out the gauge fields $A_{\pm}$ the non-Abelian T-dual of the PS superstring is constructed. The content of the lambda deformation is now clear, it is a deformation of the non-Abelian T-dual\footnote{This is the main characteristic of the $\lambda$ deformed sigma models.} of the PS superstring, see \eqref{eff act} and \eqref{eff act 2} for the fully deformed effective action functional obtained after elimination of the gauge fields. As we shall see along this text, it also preserves all the properties of the un-deformed theory.

In order to avoid clutter in the expressions to be written in the rest of the paper, it will be useful to introduce the following notation, i.e.
\begin{equation}
s=\lambda^{-4}-1, \label{s}
\end{equation}
and\footnote{We have that $D(*)\equiv Ad_{\mathcal{F}}(*)=\mathcal{F}(*)\mathcal{F}^{-1}$ and $D^{T}(*)\equiv Ad_{\mathcal{F}^{-1}}(*)=\mathcal{F}^{-1}(*)\mathcal{F}$. } 
\begin{equation}
\mathcal{O}=\Omega-D. \label{O}
\end{equation} 

\subsection{Equations of motion}
Before considering the equations of motion in detail, let us first prove an useful identity. Consider the gauge field equations of
motion derived from the action \eqref{deformed PS}, i.e.%
\begin{equation}
\begin{aligned}
A_{+} &=\mathcal{O}^{-T}[\mathcal{F}^{-1}\partial _{+}\mathcal{F}%
-sN^{(0)}_{+}], \\
A_{-} &=-\mathcal{O}^{-1}[\partial _{-}\mathcal{FF}^{-1}+s%
\hat{N}^{(0)}_{-}]. \label{A eom}
\end{aligned}
\end{equation}%
Then, the Maurer-Cartan identity for the flat current $\mathcal{F}^{-1}\partial
_{\pm }\mathcal{F}$, in the presence of the equations \eqref{A eom}, takes the form 
\begin{equation}
\xi _{1}-D^{T}\xi _{2}=0,
\end{equation}%
where%
\begin{equation}
\begin{aligned}
\xi _{1} &=[\partial _{+}+\Omega ^{T}A_{+}+sN^{(0)}_{+},\partial
_{-}+A_{-}], \\
\xi _{2} &=[\partial _{+}+A_{+},\partial _{-}+\Omega A_{-}+s%
\hat{N}^{(0)}_{-}].
\end{aligned}
\end{equation}
Now, the $\mathcal{F}$ equations of motion when combined with \eqref{A eom} are equivalent to having $\xi _{1}=0$ and $\xi _{2}=0$ separately, while the ghosts equations of motion
\begin{equation}
\begin{aligned}
D_{-}^{(0)}l^{(1)}-[\hat{N}^{(0)}_{-},l^{(1)}]&=0,\text{ \ \ }D_{-}^{(0)}w^{(3)}_{+}-[\hat{N}^{(0)}_{-},w^{(3)}_{+}]=0,\\
D_{+}^{(0)}\hat{l}^{(3)}-[N^{(0)}_{+},\hat{l}^{(3)}]&=0,\text{ \ \ }D_{+}^{(0)}\hat{w}^{(1)}_{-}-[N^{(0)}_{+},\hat{w}^{(1)}_{-}]=0\label{l,w eom}
\end{aligned}%
\end{equation}
imply that the PS Lorentz currents \eqref{PS Lorentz} satisfy
\begin{equation}
D_{-}^{(0)}N^{(0)}_{+}-[\hat{N}^{(0)}_{-},N^{(0)}_{+}]=0,\text{ \ \ }D_{+}^{(0)}\hat{N}^{(0)}_{-}-[N^{(0)}_{+},%
\hat{N}^{(0)}_{-}]=0. \label{ghost eom}
\end{equation}%

In terms of the dual currents 
\EQ{
I_{\pm }^{(0)} &=A_{\pm }^{(0)},\text{ \ \ }I_{+}^{(1)}=\lambda
^{-1/2}A_{+}^{(1)},\text{ \ \ }I_{-}^{(1)}=\lambda ^{-3/2}A_{-}^{(1)}, \\
I_{\pm }^{(2)} &=\lambda ^{-1}A_{\pm }^{(2)},\text{ \ \ }%
I_{+}^{(3)}=\lambda ^{-3/2}A_{+}^{(3)},\text{ \ \ }I_{-}^{(3)}=\lambda
^{-1/2}A_{-}^{(3)}, 
\label{Dual currents} 
}%
introduced in \cite{exploring} for the lambda deformed hybrid formulation of the superstring, the $\mathcal{F}$ equations of motion, i.e. $\xi _{1}=\xi _{2}=0$ are (for generic values of $\lambda$) equivalent to the following set of equations
\begin{equation}
\begin{aligned}
D_{+}^{(0)}I_{-}^{(3)}+[I_{+}^{(1)},I_{-}^{(2)}]+[I_{+}^{(2)},I_{-}^{(1)}]-[N^{(0)}_{+},I_{-}^{(3)}]+\lambda^{-2}[I_{+}^{(3)},\hat{N}^{(0)}_{-}]
&=0,   \\
D_{-}^{(0)}I_{+}^{(1)}+[I_{-}^{(3)},I_{+}^{(2)}]+[I_{-}^{(2)},I_{+}^{(3)}]-[\hat{N}^{(0)}_{-}, I_{+}^{(1)}]+\lambda^{-2}[I_{-}^{(1)},N^{(0)}_{+}]
&=0,   \\
D_{+}^{(0)}I_{-}^{(2)}+[I_{+}^{(1)},I_{-}^{(1)}]-[N^{(0)}_{+},I_{-}^{(2)}]+\lambda^{-2}[I_{+}^{(2)},\hat{N}^{(0)}_{-}] &=0,\text{ } \\
\text{\ }D_{-}^{(0)}I_{+}^{(2)}+[I_{-}^{(3)},I_{+}^{(3)}]-[\hat{N}^{(0)}_{-},I_{+}^{(2)}]+\lambda^{-2}[I_{-}^{(2)},N^{(0)}_{+}]&=0,   \\
D_{+}^{(0)}I_{-}^{(1)}-[N^{(0)}_{+},I_{-}^{(1)}]+\lambda^{-2}[I_{+}^{(1)},\hat{N}^{(0)}_{-}] &=0,\text{ \ }   \\
D_{-}^{(0)}I_{+}^{(3)}-[\hat{N}^{(0)}_{-},I_{+}^{(3)}]+\lambda^{-2}[I_{-}^{(3)},N^{(0)}_{+}] &=0, \\ 
F_{+-}^{(0)}+s[N^{(0)}_{+},\hat{N}^{(0)}_{-}] &=0,
\label{k-def eom} 
\end{aligned}
\end{equation}
where
\begin{equation}
F_{+-}^{(0)}=\partial _{+}I_{-}^{(0)}-\partial_{-}I_{+}^{(0)}+[I_{+}^{(0)},I_{-}^{(0)}]+[I_{+}^{(1)},I_{-}^{(3)}]+[I_{+}^{(2)},I_{-}^{(2)}]+[I_{+}^{(3)},I_{-}^{(1)}].
\end{equation}

The full set of equations of motion \eqref{ghost eom} and \eqref{k-def eom} follow from the zero curvature condition of the Lax connection
\begin{equation}
\begin{aligned}
\mathscr{L}_{+}(z)
&=I_{+}^{(0)}+zI_{+}^{(1)}+z^{2}I_{+}^{(2)}+z^{3}I_{+}^{(3)}+(z^{4}-\lambda
^{2})\lambda ^{-2} N^{(0)}_{+}, \\
\mathscr{L}_{-}(z)
&=I_{-}^{(0)}+z^{-3}I_{-}^{(1)}+z^{-2}I_{-}^{(2)}+z^{-1}I_{-}^{(3)}+(z^{-4}-\lambda ^{2})\lambda^{-2} \hat{N}^{(0)}_{-}, \label{Lax on}
\end{aligned}
\end{equation}
satisfying the condition
\begin{equation}
\Phi (\mathscr{L}_{\pm}(z))=\mathscr{L}_{\pm}(iz), \label{phi on Lax}
\end{equation}
under the action of $\Phi$ in \eqref{auto}.

However, in contrast to all known lambda models, see for instance \cite{Sfetsos,lambda-bos,lambda-fer,exploring}, this theory has an explicit $\lambda$-dependent Lax connection, i.e. the parameter $\lambda$ can not be absorbed by the ghost currents. Indeed, the un-deformed theory has a Lax pair given by \cite{breno}
\begin{equation}
\begin{aligned}
\mathscr{L}_{+}(z)
&=J_{+}^{(0)}+zJ_{+}^{(1)}+z^{2}J_{+}^{(2)}+z^{3}J_{+}^{(3)}+(z^{4}-1) N^{(0)}_{+}, \\
\mathscr{L}_{-}(z)
&=J_{-}^{(0)}+z^{-3}J_{-}^{(1)}+z^{-2}J_{-}^{(2)}+z^{-1}J_{-}^{(3)}+(z^{-4}-1) \hat{N}^{(0)}_{-}, \label{Lax Undef}
\end{aligned}
\end{equation}
where $J_{\pm}=f^{-1}\partial_{\pm}f$ is a flat current defined in terms of the Lagrangian field $f$. The integrability of the action \eqref{deformed PS} was not considered in \cite{exploring} because of the discrepancy of the equations of motion \eqref{k-def eom} with the equations of motion of the un-deformed theory. Notice it explicit $\lambda$-dependence. However, this apparent anomalous behavior is quite natural once we realize it is just a consequence of the pole structure of the deformed theory, materialized in the twisting function, see \eqref{twisting function} below.

Using the Kac-Moody currents expressions defined below in \eqref{KM currents off}, we can write \eqref{A eom} in the equivalent forms%
\begin{equation}
\begin{aligned}
\frac{2\pi }{k}\mathscr{J}_{+} &=\Omega ^{T}A_{+}-A_{-}+sN^{(0)}_{+},
\\
-\frac{2\pi }{k}\mathscr{J}_{-} &=A_{+}-\Omega A_{-}-s%
\hat{N}^{(0)}_{-},
\end{aligned}
\end{equation}%
from which follows that%
\begin{equation}
\mathscr{L}_{\sigma }(z_{\pm })=\mp \frac{2\pi }{k}\mathscr{J}_{\mp }, \label{Lax at poles}
\end{equation}%
where $z_{\pm }=\lambda ^{\pm 1/2}.$ This important result will be invoked later on. 

The lambda deformation preserves the integrability of the original theory albeit with a slight modification of the Lax connection when compared to its un-deformed counterpart.

\subsection{BRST symmetry}

The action \eqref{deformed PS} is invariant under the following BRST variations \cite{exploring}
\begin{equation}
\begin{aligned}
\overline{\delta }\mathcal{F}&=-\alpha \mathcal{F}+\mathcal{F}\beta ,\text{
\ \ \ \ }\overline{\delta }\hat{w}^{(1)}_{-}=-\lambda bA_{-}^{(1)}, \\
\overline{\delta }A_{+} &=D_{+}\alpha ,\text{\; \; \ \ \ \ \ \ \ \ \ \ }\overline{%
\delta }w^{(3)}_{+}=-\lambda aA_{+}^{(3)}, \\
\overline{\delta }A_{-} &=D_{-}\beta ,\text{ \; \; \; \; \; \; \; \ }\overline{%
\delta }l^{(1)}=\overline{\delta }\hat{l}^{(3)}=0,\text{\ \ \ } \label{BRST off}
\end{aligned}
\end{equation}%
where $a,b\in 
\mathbb{R}
$ are arbitrary real numbers\footnote{This is a freedom found in \cite{exploring}, where the BRST symmetry variations were constructed.} and%
\begin{equation}
\alpha =\lambda al^{(1)}+b\hat{l}^{(3)},\text{ \ \ }\beta
=al^{(1)}+\lambda b\hat{l}^{(3)}. \label{alpha,beta}
\end{equation}
We can verify the consistency of the BRST variations by showing that its square is formally a gauge transformation. We find that
\begin{equation}
\begin{aligned}
\overline{\delta }^{2}\mathcal{F}&=-[\alpha^{2}, \mathcal{F}],\text{
\ \ \ \ }\overline{\delta }^{2}\hat{w}^{(1)}_{-}=-\lambda b(D_{-}\beta)^{(1)}, \\
\overline{\delta }^{2}A_{+} &=D_{+}\alpha^{2} ,\text{ \; \ \ \ \ \ \ }\overline{%
\delta }^{2}w^{(3)}_{+}=-\lambda a(D_{+}\alpha)^{(3)}, \\
\overline{\delta }^{2}A_{-} &=D_{-}\alpha^{2} ,\text{ \; \; \. \; \; \ }\overline{%
\delta }^{2}l^{(1)}=\overline{\delta}^{2} \hat{l}^{(3)}=0,\text{\ \ \ } \label{square BRST}
\end{aligned}
\end{equation}%
where we have used the constraints \eqref{PS constraints} in order to show that $\alpha^{2}=\beta^{2}$. Classical nilpotency of the BRST action must be accomplished up to classical equations of motion and local gauge transformations. In relation to the action functional \eqref{deformed PS}, we find that such an action is indeed nilpotent
\begin{equation}
\overline{\delta }^{2}S_{\lambda}=0.
\end{equation}
In showing this last result, derivatives of the constraints \eqref{PS constraints} are to be used. This is consistent with the bosonic gauge symmetry of \eqref{deformed PS} generated by the grade zero subalgebra $\mathfrak{f}^{(0)}$. In particular, \eqref{square BRST} shows that for the lambda deformation of the PS superstring the
Lorentz transformation must be modified as 
\begin{equation}
\delta_{Lor} \mathcal{F}=[\Lambda^{(0)},\mathcal{F}],
\end{equation}
as expected for a lambda deformation. Recall that in the un-deformed case the Lorentz transformation is of the form
\begin{equation}
\delta_{Lor} f=\Lambda^{(0)}f.
\end{equation} 

In order to find the associated BRST charge in an elegant manner, we consider the symplectic form of the action \eqref{deformed PS}. Namely,
\begin{equation}
\omega =\omega _{+}-\omega _{-},
\end{equation}%
where%
\begin{equation}
\begin{aligned}
\omega _{+} &=\frac{k}{4\pi }\int d\sigma ^{+}\langle \delta \mathcal{%
FF}^{-1}\wedge D_{+}(\delta \mathcal{FF}^{-1})+2\delta \mathcal{FF}%
^{-1}\wedge \delta A_{+}+2(\lambda ^{-4}-1)\delta w_{+}^{(3)}\wedge \delta
l^{(1)}\rangle , \\
\omega _{-} &=\frac{k}{4\pi }\int d\sigma ^{-}\langle \mathcal{F}%
^{-1}\delta \mathcal{F}\wedge D_{-}(\mathcal{F}^{-1}\delta \mathcal{F})-2%
\mathcal{F}^{-1}\delta \mathcal{F}\wedge \delta A_{-}+2(\lambda
^{-4}-1)\delta \hat{w}_{-}^{(1)}\wedge \delta \hat{l}%
^{(3)}\rangle. 
\end{aligned}
\end{equation}%
Here, the $\delta$ is to be understood as the exterior derivative in the symplectic manifold. Using the following contractions%
\begin{equation}
\begin{aligned}
\delta \mathcal{F}(X)&=-\alpha \mathcal{F}+\mathcal{F}\beta ,\text{
\ \ \ \ }\delta \hat{w}^{(1)}_{-}(X)=-\lambda bA_{-}^{(1)}, \\
\delta A_{+}(X) &=D_{+}\alpha ,\text{\; \; \ \ \ \ \ \ \ \ \ \ }
\delta w^{(3)}_{+}(X)=-\lambda aA_{+}^{(3)}, \\
\delta A_{-}(X) &=D_{-}\beta ,\text{ \; \; \; \; \; \; \; \ }
\delta l^{(1)}(X)=\delta \hat{l}^{(3)}(X)=0,\text{\ \ \ }
\end{aligned}
\end{equation}%
we find that%
\begin{equation}
-i_{X}\omega =\delta Q_{BRST},
\end{equation}%
where%
\begin{equation}
Q_{BRST}=-\frac{k}{2\pi }\frac{s}{\lambda}\int\nolimits_{S^{1}}d\sigma \langle al^{(1)}A_{+}^{(3)}+b%
\hat{l}^{(3)}A_{-}^{(1)}\rangle . \label{BRST charge}
\end{equation}%
This result is found only after using the gauge field equations of motion \eqref{A eom},
which allow to write the right hand side as a total differential. 

We now consider the conservation and the nilpotency of \eqref{BRST charge}. Using \eqref{PS constraints}, \eqref{Dual currents} and the equations of motion \eqref{k-def eom}, we get
\begin{equation}
\partial_{\tau}Q_{BRST}=-\frac{k}{2\pi }\frac{s}{\lambda}\int\nolimits_{S^{1}}d\sigma \partial_{\sigma} \langle al^{(1)}A_{+}^{(3)}-b%
\hat{l}^{(3)}A_{-}^{(1)}\rangle, 
\end{equation}%
ensuring the conservation of the BRST charge after imposing periodic boundary conditions on the fields. Concerning the nilpotency of the BRST charge care must taken because of the equations \eqref{A eom} imply that $A_{\pm}$ are now determined by the Lagrangian field $\mathcal{F}$ and consistency of the BRST symmetry must be guaranteed first. This means that the correct variations $\overline{\delta }A_{\pm}$ must be deduced from the variation\footnote{Recall that in \eqref{BRST off}, $A_{\pm}$ and $\mathcal{F}$ are taken as independent fields.} $\overline{\delta }\mathcal{F}$. In order to show this, we start with the general result   
\begin{equation}
\begin{aligned}
\overline{\delta }A_{+} &=\mathcal{O}^{-T}\{ D^{T}D_{+}(\overline{%
\delta }\mathcal{FF}^{-1})-s\overline{\delta }N_{+}^{(0)}\} , \\
\overline{\delta }A_{-} &=-\mathcal{O}^{-1}\{ DD_{-}(\mathcal{F}^{-1}%
\overline{\delta }\mathcal{F})+s\overline{\delta }\hat{N}%
_{-}^{(0)}\} 
\end{aligned}
\end{equation}
and use the following BRST variations
\begin{equation}
\begin{aligned}
\overline{\delta }\mathcal{F}&=-\alpha \mathcal{F}+\mathcal{F}\beta ,\text{
\ \ \ \ }\overline{\delta }\hat{w}^{(1)}_{-}=-\lambda bA_{-}^{(1)}, \\
\overline{\delta }w^{(3)}_{+}&=-\lambda aA_{+}^{(3)},\text{
\ \ \ \ \ \ \ \ \ } \overline{\delta }l^{(1)}=\overline{\delta }\hat{l}^{(3)}=0, \label{on-shell BRST}
\end{aligned}
\end{equation}%
with $A_{-}^{(1)}$ and $A_{+}^{(3)}$ extracted from \eqref{A eom}. After imposing the PS constraints \eqref{PS constraints}, we obtain
\begin{equation}
\begin{aligned}
\overline{\delta }A_{+} &=D_{+}\alpha -s\lambda b\mathcal{O}^{-T}\{
D_{+}^{(0)}\hat{l}^{(3)}-[N_{+}^{(0)},\hat{l}^{(3)}]\} , \\
\overline{\delta }A_{-} &=D_{-}\beta -s\lambda a\mathcal{O}^{-1}\{
D_{-}^{(0)}l^{(1)}-[\hat{N}_{-}^{(0)},l^{(1)}]\}. 
\end{aligned}
\end{equation}
This shows that the BRST symmetry is consistent with \eqref{A eom} only after the use of PS ghosts equations of motion \eqref{l,w eom}, i.e. the BRST symmetry is now an on-shell symmetry. This same situation occurs in the un-deformed theory. Now, using the new BRST symmetry transformations \eqref{on-shell BRST} together with \eqref{PS constraints}, we verify that the BRST charge is indeed nilpotent
\begin{equation}
\overline{\delta }Q_{BRST}=0. 
\end{equation}%
Furthermore, it follows from the ghost equations of motion, \eqref{k-def eom} and \eqref{PS constraints} that the currents defined by
\begin{equation}
j_{+}(\sigma^{+})=\langle l^{(1)}A_{+}^{(3)}\rangle ,\text{ \ \ }%
j_{-}(\sigma^{-})=\langle \hat{l}^{(3)}A_{-}^{(1)}\rangle, \label{chiral BRST}
\end{equation}%
are chiral, just as their counterparts in the un-deformed theory. A comment is in order, it is known \cite{E-models} that the eta and the lambda backgrounds are related via Poisson-Lie T-duality plus analytic continuation of their deformation parameters (at least for purely bosonic theories), thus it is sensible to think that the charge \eqref{BRST charge} and the BRST charge of the eta deformed pure spinor supertring \cite{eta-PS} might be related under this type of duality.

In summary, all the properties of the original action functional are preserved under the deformation, i.e. its BRST symmetry, its integrability and its local symmetries. Below, we will show that its 1-loop conformal symmetry is also maintained.

\section{Hamiltonian structure and integrability}\label{3}

In this section, we run the Dirac procedure and study the integrable structure from the Hamiltonian theory point of view. We will follow the strategy of \cite{Magro,Vicedo,lambdaCS,lambdaCS2} and show that the Poisson bracket of the spatial component of the extended Lax connection takes the Maillet algebra form \cite{Maillet}. This is possible after constructing a suitable extension of the Lax connection outside the constraint surface. As expected, the extended monodromy matrix is conserved and their charges preserve the constraint surface where the classical motion of the deformed string theory takes place.

\subsection{Dirac procedure}

The phase space associated to the action functional \eqref{deformed PS} is described by the following phase space coordinates: two currents $\mathscr{J}_{\pm }$ given by
\begin{equation}
\mathscr{J}_{+}=\frac{k}{2\pi }\left( \mathcal{F}^{-1}\partial _{+}\mathcal{%
F+F}^{-1}A_{+}\mathcal{F-}A_{-}\right) ,\text{ \ \ }\mathscr{J}_{-}=-\frac{k}{%
2\pi }\left( \partial _{-}\mathcal{F\mathcal{F}}^{-1}\mathcal{-F}A_{-}%
\mathcal{F}^{-1}\mathcal{+}A_{+}\right),  \label{KM currents off}
\end{equation}%
obeying the relations of two commuting Kac-Moody algebras\footnote{For the Lie (super)-algebra we use the definitions
$\eta _{AB}=\left\langle T_{A},T_{B}\right\rangle ,$ $%
C_{\mathbf{12}}$ = $\eta ^{AB}T_{A}\otimes T_{B}$ and $u_{\mathbf{1}}=u\otimes I,$ $u_{\mathbf{2}}%
=I\otimes u$, etc.}
\begin{equation}
\{\mathscr{J}_{\pm }(\sigma)_{\mathbf{1}},\mathscr{J}_{\pm }(\sigma^{\prime})_{\mathbf{2}}\}=%
-[C_{\mathbf{12}},\mathscr{J}%
_{\pm }(\sigma^{\prime})_{\mathbf{2}}]\delta _{\sigma\sigma^{\prime}}\mp \frac{k}{2\pi }C_{\mathbf{12}}\delta _{\sigma\sigma^{\prime}}^{\prime },%
\text{ \ \ }\{\mathscr{J}_{\pm }(\sigma)_{\mathbf{1}},\mathscr{J}%
_{\mp }(\sigma^{\prime})_{\mathbf{2}}\}=0,  \label{super KM algebra}
\end{equation}%
two conjugated  pairs of fields $(A_{\pm},P_{\mp})$ with Poisson brackets
\begin{equation}
\{P_{\pm }(\sigma)_{\mathbf{1}},A_{\mp }(\sigma^{\prime})_{\mathbf{2}}\}=\frac{1}{2}C_{\mathbf{12}}\delta _{\sigma\sigma^{\prime}} \label{PS brackets}
\end{equation}
and two pairs of conjugated ghosts $(l^{(1)},w^{(3)}_{+})$ and $(\hat{l}^{(3)},\hat{w}^{(1)}_{-})$, satisfying
\begin{equation}
\{l^{(1)}(\sigma )_{\mathbf{1}},w^{(3)}_{+}(\sigma ^{\prime })_{\mathbf{2}%
}\}=-\alpha \lambda ^{2}C_{\mathbf{12}}^{(13)}\delta _{\sigma\sigma^{\prime}},\text{ \ \ }\{\hat{l}^{(3)}(\sigma )_{\mathbf{1}},%
\hat{w}^{(1)}_{-}(\sigma ^{\prime })_{\mathbf{2}}\}=-\alpha \lambda ^{2}C_{%
\mathbf{12}}^{(31)}\delta _{\sigma\sigma^{\prime}},
\end{equation}
where\footnote{Not to be confused with the alpha defined in \eqref{alpha,beta}.}
\begin{equation}
\alpha=\frac{2\pi}{k}\frac{1}{(\lambda^{-2}-\lambda^{2})}.
\end{equation}
The time flow is determined by the canonical Hamiltonian density
\begin{equation}
H_{C}=H_{\Omega}+H_{PS}, \label{canonical H}
\end{equation}
where
\begin{equation}
H_{\Omega}=-\frac{k}{\pi }\big\langle \left( \frac{\pi }{k}\right) ^{2}\left( 
\mathscr{J}_{+}^{2}+\mathscr{J}_{-}^{2}\right) +\frac{2\pi }{k}\left( A_{+}%
\mathscr{J}_{-}+A_{-}\mathscr{J}_{+}\right) +\frac{1}{2}\left(
A_{+}^{2}+A_{-}^{2}\right) -A_{+}\Omega A_{-}\big\rangle 
\end{equation}%
and
\begin{equation}
H_{PS}=-\frac{1}{\alpha \lambda^{2}}\big\langle w^{(3)}_{+}\partial
_{\sigma }l^{(1)}-\hat{w}^{(1)}_{-}\partial _{\sigma }\hat{l}%
^{(3)}-2N^{(0)}_{+}A_{-}-2\hat{N}^{(0)}_{-}A_{+}+2N^{(0)}_{+}\hat{N}^{(0)}_{-}\big\rangle 
\end{equation}
through the relation 
\begin{equation}
\partial _{\tau }f =\big\{ f ,h_{C}\big\} ,%
\text{ \ \ }h_{C}=\int\nolimits_{S^{1}}d\sigma H_{C}(\sigma
).
\end{equation}
Above, $f$ is an arbitrary functional of the phase space variables.

Now, we run the Dirac algorithm. There are two primary constraints
\begin{equation}
P_{+}\approx 0,\text{ \ \ }P_{-}\approx 0. \label{primary const}
\end{equation}%
By adding them to the canonical Hamiltonian \eqref{canonical H} we construct the total Hamiltonian
\begin{equation}
H_{T}=H_{C}-2\left\langle u_{+}P_{-}+u_{-}P_{+}\right\rangle,
\end{equation}%
where $u_{\pm}$ are arbitrary Lagrange multipliers. 

Stability of the primary constraints \eqref{primary const} under the flow of $H_{T}$ leads to two secondary constraints given by
\begin{equation}
\begin{aligned}
C_{+}&=\mathscr{J}_{+}-\frac{k}{2\pi }\big( \Omega ^{T}A_{+}-A_{-}+sN^{(0)}_{+}\big)
\approx 0, \\
C_{-}&=\mathscr{J}_{-}+\frac{k}{2\pi }\big(A_{+}-\Omega
A_{-}-s\hat{N}^{(0)}_{-}\big) \approx 0,  \label{secondary const}
\end{aligned}
\end{equation}%
which are nothing but the gauge field equations of motion \eqref{A eom}.
In this formulation of the superstring we must add by hand the pure spinor constraints \eqref{PS constraints} to the set of constraints found so far, i.e.
\begin{equation}
\Phi =\frac{1}{2}l^{(1)}l^{(1)}\approx 0,\text{ \ \ }\hat{\Phi }=\frac{1%
}{2}\hat{l}^{(3)}\hat{l}^{(3)}\approx 0. \label{PS constraints 2}
\end{equation}
From this, we construct the extended Hamiltonian 
\begin{equation}
H_{E}=H_{C}-2\big\langle u_{+}P_{-}+u_{-}P_{+}+\mu_{+}C_{-}+\mu_{-}C_{+}+v\Phi+\hat{v}\hat{\Phi}\big\rangle,
\end{equation}
where $\mu_{\pm}$ and $v,\hat{v}$ are arbitrary Lagrange multipliers. 

Stability of the constraints under the time flow of $H_{E}$ produce no new constraints but rather determine some of the Lagrange multipliers. However, their explicit form will not be required in what follows and algorithm stops.

We now consider the constraints and split them between first and second class. Along the coset directions, the constraints 
\begin{equation}
P_{\pm}^{(i)}\approx 0,\text{ \ \ }C_{\pm }^{(i)}\approx 0,\text{ \ \ for \ \ }%
i=1,2,3 \label{const 1}
\end{equation}%
form three second class pairs of constraints and we impose them strongly by
means of a Dirac bracket. The brackets \eqref{super KM algebra} and \eqref{PS brackets} are not modified in this process, so we continue using their usual definitions. Then, we have the strong relations
\begin{equation}
\begin{aligned}
I_{+}^{(3)} &=\alpha (\lambda ^{-1/2}\mathscr{J}_{+}^{(3)}+\lambda ^{1/2}%
\mathscr{J}_{-}^{(3)}),\text{ \ \ }I_{-}^{(3)}=\alpha (\lambda ^{3/2}%
\mathscr{J}_{+}^{(3)}+\lambda ^{-3/2}\mathscr{J}_{-}^{(3)}), \\
I_{+}^{(2)} &=\alpha (\lambda ^{-1}\mathscr{J}_{+}^{(2)}+\lambda \mathscr{J}%
_{-}^{(2)}),\text{ \: \ \ \ \ \ \ }I_{-}^{(2)}=\alpha (\lambda \mathscr{J}%
_{+}^{(2)}+\lambda ^{-1}\mathscr{J}_{-}^{(2)}), \\
I_{+}^{(1)} &=\alpha (\lambda ^{-3/2}\mathscr{J}_{+}^{(1)}+\lambda ^{3/2}%
\mathscr{J}_{-}^{(1)}),\text{ \ \ }I_{-}^{(1)}=\alpha (\lambda ^{1/2}%
\mathscr{J}_{+}^{(1)}+\lambda ^{-1/2}\mathscr{J}_{-}^{(1)}). \label{I as Js}
\end{aligned}
\end{equation}

Along the grade zero part of the algebra, we notice that the combination
\begin{equation}
P_{+}^{(0)}+P_{-}^{(0)}\approx 0 \label{1st P}
\end{equation}
is a first class constraint, while 
\begin{equation}
P_{+}^{(0)}-P_{-}^{(0)}\approx 0,\text{ \ \ }C_{-}^{(0)}\approx 0 \label{2nd P}
\end{equation}
form a pair of second class constraints. The first class constraint \eqref{1st P} can be gauge fixed by means of the condition
\begin{equation}
A_{-}^{(0)}\approx 0. \label{A=0}
\end{equation}
This is a good gauge fixing condition and now we impose \eqref {1st P}, \eqref{2nd P} and \eqref{A=0} strongly by means of a Dirac bracket. Fortunately, the brackets \eqref{super KM algebra} and \eqref{PS brackets} are not modified at this step. Then, we get the strong relation
\begin{equation}
I_{1}^{(0)}=-\frac{2\pi }{k}\mathscr{J}_{-}^{(0)}+s\hat{N}^{(0)}_{-}. \label{I0 as J0}
\end{equation}

At this level of analysis, the remaining constraints are $(\varphi ^{(0)},\Phi, \hat{\Phi})$, where
\begin{equation}
\varphi ^{(0)}\equiv C_{+}^{(0)}=\mathscr{J}_{+}^{(0)}+\mathscr{J}_{-}^{(0)}-%
\frac{1}{\alpha \lambda^{2} }(N^{(0)}_{+}+\hat{N}^{(0)}_{-}).
\end{equation}
They weakly commute among themselves and are first class. 

When compared to the Hamiltonian analysis of the lambda deformed GS superstring \cite{lambdaCS2}, we realize that for the PS formalism the same analysis is simpler and less involved because of the absence of the fermionic constraints associated to the kappa symmetry. 

\subsection{Maillet algebra}

Due to the presence of Hamiltonian constraints, the Poisson bracket of the spatial component of the Lax connection does not take the standard Maillet form \cite{Maillet} and an extension of the Lax connection \eqref{Lax on} outside the constraint surface must be considered. Only after a proper extension have been chosen, the Maillet algebra is recovered. See \cite{Magro,Vicedo} for string theories with this characteristic, see also \cite{lambdaCS2} for a more direct approach devoted to lambda models. Here, we will apply the same strategy to the PS superstring lambda model.

We start by writing the spatial component of the Lax connection \eqref{Lax on} in terms of the Kac-Moody currents and in the partial gauge considered so far, where \eqref{I as Js}, \eqref{A=0} and \eqref{I0 as J0} are valid in the strong sense. We have
\begin{equation}
\begin{aligned}
\mathscr{L}_{\sigma }(z) =&-\frac{2\pi }{k}\mathscr{J}_{-}^{(0)}+f_{-}(z)%
\left\{ \frac{z_{+}^{3}}{z^{3}}\mathscr{J}_{-}^{(1)}+\frac{z_{+}^{2}}{z^{2}}%
\mathscr{J}_{-}^{(2)}+\frac{z_{+}}{z}\mathscr{J}_{-}^{(3)}\right\}  \\
&+f_{+}(z)\left\{ \frac{z_{-}^{3}}{z^{3}}\mathscr{J}_{+}^{(1)}+\frac{%
z_{-}^{2}}{z^{2}}\mathscr{J}_{+}^{(2)}+\frac{z_{-}}{z}\mathscr{J}%
_{+}^{(3)}\right\} +\frac{f_{+}(z)}{\alpha z_{+}^{4}}\left\{ N^{(0)}_{+}+\frac{%
z_{-}^{4}}{z^{4}}\hat{N}^{(0)}_{-}\right\} ,
\end{aligned}
\end{equation}%
where
\begin{equation}
f_{\pm }(z)=\alpha (z^{4}-z_{\pm }^{4}).
\end{equation}

The extension of the Lax connection we will consider is defined by imposing that the relations \eqref{Lax at poles} are valid outside
the constraint surface and hence on the whole phase space. We consider the following obvious extension\footnote{This choice is based on the extension constructed in \cite{lambdaCS2} for the GS superstring simply by dropping the contributions coming from kappa symmetry. A similar choice is considered in \cite{Magro} for the un-deformed PS superstring.} 
\begin{equation}
\overline{\mathscr{L}}_{\sigma }(z)=\mathscr{L}_{\sigma }(z)+f_{+}(z)\varphi
^{(0)} \label{Ext spat Lax}
\end{equation}%
and obtain
\begin{equation}
\overline{\mathscr{L}}_{\sigma }(z)=f_{+}(z)\Omega (z/z_{-})\mathscr{J}%
_{+}+f_{-}(z)\Omega (z/z_{+})\mathscr{J}_{-}-\frac{2\varphi _{\lambda }^{-1}(z)%
}{\alpha z_{+}^{4}}\hat{N}^{(0)}_{-}, \label{extended space}
\end{equation}%
where%
\begin{equation}
\varphi _{\lambda }^{-1}(z)=\frac{f_{+}(z)f_{-}(z)}{2\alpha z^{4}}.
\end{equation}

The Poisson bracket of \eqref{extended space} with itself takes the $\mathfrak{r}/\mathfrak{s}$ Maillet algebra form \cite{Maillet}
\begin{equation}
\begin{aligned}
\{\overline{\mathscr{L}}_{\sigma}(\sigma;z)_{\mathbf{1}},\overline{\mathscr{L}}_{\sigma}(\sigma^{\prime};w)_{\mathbf{2}}\}=[&\mathfrak{r}
_{\mathbf{12}}(z,w),\overline{\mathscr{L}}_{\sigma}(\sigma;z)_{\mathbf{1}}+\overline{\mathscr{L}}_{\sigma}(\sigma^{\prime};w)_{\mathbf{2}}]\delta
_{\sigma \sigma^{\prime}} \\
+&[\mathfrak{s}_{\mathbf{12}}(z,w),\overline{\mathscr{L}}_{\sigma}(\sigma;z)_{\mathbf{1}}-\overline{\mathscr{L}}_{\sigma}(\sigma^{\prime};w)_{\mathbf{2}}]\delta _{\sigma \sigma^{\prime}}-2\mathfrak{s}_{\mathbf{12}}(z,w)\delta _{\sigma \sigma^{\prime}}^{\prime },
\label{Maillet-lambda}
\end{aligned}
\end{equation}%
where
\begin{equation}
\begin{aligned}
\mathfrak{r}_{\mathbf{12}}(z,w)&=-\frac{1}{z^{4}-w^{4}}\tsum\nolimits_{j=0}^{3}%
\{z^{j}w^{4-j}C_{\mathbf{12}}^{(j,4-j)}\varphi _{\lambda
}^{-1}(w)+z^{4-j}w^{j}C_{\mathbf{12}}^{(4-j,j)}\varphi _{\lambda }^{-1}(z)\}, \\
\mathfrak{s}_{\mathbf{12}}(z,w)&=-\frac{1}{z^{4}-w^{4}}\tsum\nolimits_{j=0}^{3}%
\{z^{j}w^{4-j}C_{\mathbf{12}}^{(j,4-j)}\varphi _{\lambda
}^{-1}(w)-z^{4-j}w^{j}C_{\mathbf{12}}^{(4-j,j)}\varphi _{\lambda }^{-1}(z)\}, \label{def r,s}
\end{aligned}
\end{equation}%
are the anti-symmetric and symmetric parts of an $R_{\mathbf{12}}(z,w)$ matrix and $\varphi _{\lambda }(z)$ is the associated deformed twisting function given by
\begin{equation}
\varphi _{\lambda }(z)=\frac{2}{\alpha }.\frac{1}{%
(z^{2}-z^{-2})^{2}-(z_{+}^{2}-z_{-}^{2})^{2}}. \label{twisting function}
\end{equation}%
The algebra \eqref{Maillet-lambda} is the same found for the Green-Schwarz \cite{lambdaCS2} and the hybrid formulations \cite{exploring}. At the points $z=z_{\pm}$, the Maillet algebra \eqref{Maillet-lambda} reduce to the Kac-Moody algebras we wrote above in \eqref{super KM algebra}.

The extended Lax connection is given by
\begin{equation}
\overline{\mathscr{L}}_{+}(z)=\mathscr{L}_{+}(z)+f_{+}(z)\varphi ^{(0)},%
\text{ \ \ }\overline{\mathscr{L}}_{-}(z)=\mathscr{L}_{-}(z) \label{Lax off}
\end{equation}
and from it we get an extension of the equations of motion found from \eqref{Lax on} by terms involving the bosonic constraint $\varphi^{(0)}$. Furthermore, from its flatness, it follows that the time derivative of the (super)-trace of the monodromy matrix
\begin{equation}
m(z)=P\exp \big[ -\oint\nolimits_{S^{1}}d\sigma \overline{\mathscr{L}}%
_{\sigma }(\sigma ;z)\big] , \label{monodromy}
\end{equation}%
vanishes, i.e. 
\begin{equation}
\frac{d}{d\tau} \langle m(z) \rangle=0.
\end{equation}
As a consequence, an infinite number of integrals of motion can be found by expanding the monodromy matrix in powers of the spectral parameter $z$. The extended Lax connection \eqref{Lax off} satisfy the condition \eqref{phi on Lax} as well.

The Poisson brackets of \eqref{extended space} with \eqref{PS constraints 2} weakly vanish, while the Poisson bracket of \eqref{extended space} with $\varphi^{(0)}$ is a gauge transformation, in the sense that
\begin{equation}
\big\{\overline{\mathscr{L}}_{\sigma }(\sigma ;z)_{\mathbf{1}},%
\varphi ^{(0)}(\sigma ^{\prime })_{\mathbf{2}}\big\}  =-C_{\mathbf{12}}^{(00)}\delta _{\sigma
\sigma ^{\prime }}^{\prime }+\big[ C_{\mathbf{12}}^{(00)},\overline{%
\mathscr{L}}_{\sigma }(\sigma ;z)_{\mathbf{1}}\big] \delta _{\sigma \sigma ^{\prime }},
\end{equation}
meaning that the (super)-trace of the mononodromy matrix Poisson commute with $\varphi^{(0)}$
\begin{equation}
\big\{\langle m(z) \rangle, \varphi^{(0)}\big\}  =0.
\end{equation}
Then, the monodromy matrix is a first class function in phase space, preserving the constraint surface where the lambda model motion takes place.

Finally, it is not difficult to verify that the BRST chiral currents \eqref{chiral BRST} remain chiral under the extended set of equations of motion.

\section{Conformal invariance}\label{4}

In this section we consider the one loop conformal symmetry of the deformed theory by following the method of \cite{Tim-beta} used to compute the 1-loop beta function of the lambda deformed GS superstring. The same method was used in \cite{exploring} to deal with the lambda deformation of the hybrid superstring. As the PS superstring is essentially based on the hybrid formulation plus the addition of the pure spinor ghosts, the calculation is quite straightforward. We also consider the supergeometry underlying the action functional \eqref{deformed PS} and show that it is the same of the GS superstring in the $AdS_{5} \times S^{5}$ lambda background \cite{Borsato}. Thus, both theories describe the same classical theory.

\subsection{1-loop beta function: un-deformed case}

For the sake of completeness and in order to understand the method of \cite{Tim-beta} in a known situation, we will compute here the 1-loop beta function of the PS superstring in the $AdS_{5} \times S^{5}$ background. We calculate the fluctuations around the background field in terms of the currents of the theories rather than their fundamental field directly, as was originally done in \cite{breno 2}. 

The equations of motion of the un-deformed theory are obtained from the flatness condition of the Lax connection \eqref{Lax Undef}. They are given by \eqref{k-def eom} after the substitutions
\begin{equation}
I_{\pm }\rightarrow J_{\pm },\text{ \ \ }\lambda \rightarrow 1.
\end{equation}
The ghost equations of motion are still given by \eqref{ghost eom} without any modification. The approach of \cite{Tim-beta} is based on taking the variations of the equations of motion in order to obtain the operators governing the fluctuations of the currents in an straightforward way. 

By choosing a purely bosonic background, the bosonic and fermionic sectors completely decouple at the 1-loop order. We choose
\begin{equation}
I_{\pm }^{(2)}= \theta _{\pm },\text{ \ \ }N^{(0)}_{+}=n_{+},\text{ \ \ }\hat{N}^{(0)}_{-}=\hat{n}_{-},
\end{equation}%
satisfying the conditions
\begin{equation}
\lbrack \theta _{\mu },\theta _{\nu }]=[\theta _{\mu },n_{+}]=[\theta
_{\mu },\hat{n}_{-}]=[n_{+},\hat{n}_{-}]=0.
\end{equation}
Other current components being zero, meaning that at the group level this choice corresponds to 
\begin{equation}
f=\exp x^{\mu }\theta _{\mu },\text{ \ \ }\theta _{\mu }\in\mathfrak{f}^{(2)}.
\end{equation}
The equations of motion are to be supplemented with a gauge-fixing condition associated to the $\mathfrak{f}^{(0)}$ gauge symmetry. We choose the following one
\begin{equation}
\partial _{+}J_{-}^{(0)}+\partial _{-}J_{+}^{(0)}=0.
\end{equation}

After variation, we obtain the operators
governing the fluctuations. For the
bosonic sector we get
\begin{equation}
\mathcal{D}_{B}(x)=\left( 
\begin{array}{cccccc}
\partial _{-}-\hat{n}_{-} & -n_{+} & 0 & -\theta _{+} &  \theta _{-} & \theta _{+}
\\ 
-\hat{n}_{-} & \partial _{+}-n_{+} & -\theta _{-} & 0 & \theta _{-} &  \theta _{+}
\\ 
-\theta _{-} & \theta _{+} & -\partial _{-} & \partial _{+} & 0 & 0 \\ 
0 & 0 & \partial _{-} & \partial _{+} & 0 & 0 \\ 
0 & 0 & 0 & -n_{+} & \partial_{-}-\hat{n}_{-} & n_{+} \\ 
0 & 0 & -\hat{n}_{-} & 0 & \hat{n}_{-} & \partial _{+}-n_{+}%
\end{array}%
\right)  \text{ \ \ acting on \ \ }\left( 
\begin{array}{c}
\delta J_{+}^{(2)} \\ 
\delta J_{-}^{(2)} \\ 
\delta J_{+}^{(0)} \\ 
\delta J_{-}^{(0)} \\
\delta N^{(0)}_{+} \\
\delta \hat{N}^{(0)}_{-}
\end{array}%
\right) , \label{det bos }
\end{equation}%
where the fourth line from top to bottom corresponds to the variation of the gauge fixing condition. For the
fermionic sector, we obtain
\begin{equation}
\mathcal{D}_{F}(x)=\left( 
\begin{array}{cccc}
\partial _{-}-\hat{n}_{-} & - n _{+} &\theta _{-} & -\theta _{+} \\ 
- \hat{n} _{-} & \partial _{+}-n_{+} & 0 & 0 \\ 
0 & 0 & \partial _{-}-\hat{n}_{-} & - n _{+} \\ 
-\theta _{-} & \theta _{+} & - \hat{n} _{-} & \partial _{+}-n_{+}%
\end{array}%
\right) \text{ \ \ acting on \ \ }\left( 
\begin{array}{c}
\delta J_{+}^{(1)} \\ 
\delta J_{-}^{(1)} \\ 
\delta J_{+}^{(3)} \\ 
\delta J_{-}^{(3)}%
\end{array}%
\right) . \label{det fer }
\end{equation}
In \eqref{det bos } and \eqref{det fer } by acting on, we mean the adjoint action, e.g. $\theta_{+}(\ast)$ means $[\theta_{+},\ast]$ and so on.

The 1-loop contribution to the effective Lagrangian, in Euclidean signature, is
\begin{equation}
\mathcal{L}_{E}^{(1)}=\frac{1}{2}\dint\nolimits_{|p|<\mu }\frac{d^{2}p}{%
(2\pi )^{2}}tr[\log \mathcal{D}_{B}(p)-\log \mathcal{D}_{F}(p)],
\end{equation}
where
\begin{equation}
\mathcal{D}_{B}(p)=\left( 
\begin{array}{cccccc}
p_{-}-\hat{n} _{-} & -n_{+} & 0 & -\theta _{+} &  \theta _{-} & \theta _{+}
\\ 
-\hat{n} _{-} & p_{+}-n_{+} & -\theta _{-} & 0 & \theta _{-} &  \theta _{+}
\\ 
-\theta _{-} & \theta _{+} & -p_{-} & p_{+} & 0 & 0 \\ 
0 & 0 & p_{-} & p_{+} & 0 & 0 \\ 
0 & 0 & 0 & -n_{+} & p_{-}-\hat{n} _{-} & n_{+} \\ 
0 & 0 & -\hat{n} _{-} & 0 & \hat{n} _{-} & p_{+}-n_{+}%
\end{array}%
\right)
\end{equation}%
and
\begin{equation}
\mathcal{D}_{F}(p)=\left( 
\begin{array}{cccc}
p_{-}-\hat{n} _{-} & - n _{+} &\theta _{-} & -\theta _{+} \\ 
- \hat{n} _{-} & p_{+}-n_{+} & 0 & 0 \\ 
0 & 0 & p _{-}-\hat{n} _{-} & - n _{+} \\ 
-\theta _{-} & \theta _{+} & - \hat{n} _{-} & p_{+}-n_{+}%
\end{array}%
\right).
\end{equation}
The contributions associated to the logarithmic divergences (denoted by $\overset{\cdot }{=}$) are
\begin{equation}
\begin{aligned}
\frac{1}{2}\dint\nolimits_{|p|<\mu }\frac{d^{2}p}{(2\pi )^{2}}tr[\log 
\mathcal{D}_{B}(p)]&\overset{\cdot }{=}-\frac{1}{8\pi }\ln \mu \lbrack
Tr_{adj}^{(0)}+Tr_{adj}^{(2)}]\left( \theta _{+}\theta _{-}+2n_{+}\hat{n} _{-}\right) , \\
\frac{1}{2}\dint\nolimits_{|p|<\mu }\frac{d^{2}p}{(2\pi )^{2}}tr[\log 
\mathcal{D}_{F}(p)]&\overset{\cdot }{=}-\frac{1}{8\pi }\ln \mu \lbrack
Tr_{adj}^{(1)}+Tr_{adj}^{(3)}]\left( \theta _{+}\theta _{-}+2n_{+}\hat{n} _{-}\right) .
\end{aligned}
\end{equation}
Then, $\mathcal{L}^{(1)}_{E}\approx c_{2}(\mathfrak{f})=0$, matching perfectly with the known result originally obtained in \cite{breno 2}.

\subsection{1-loop beta function: deformed case}

In order to compute the 1-loop beta function of the deformed theory, we consider the following classical background fields 
\begin{equation}
\mathcal{F}=\exp x^{\mu }\Lambda _{\mu },\text{ \ \ }N^{(0)}_{+}=n_{+},\text{ \ \ }\hat{N}^{(0)}_{-}=\hat{n}_{-},
\end{equation}%
satisfying the conditions
\begin{equation}
\lbrack \Lambda _{\mu },\Lambda _{\nu }]=[\Lambda _{\mu },n_{+}]=[\Lambda
_{\mu },\hat{n}_{-}]=[n_{+},\hat{n}_{-}]=0,
\end{equation}
where $\Lambda _{\mu }\in \mathfrak{f}^{(2)}$. From this choice and \eqref{A eom}, we get the dual currents%
\begin{equation}
I_{\pm }^{(2)}\equiv \theta _{\pm }=\pm \frac{\lambda }{(1-\lambda ^{2})}%
\Lambda _{\pm },\text{ \ \ }I_{\pm }^{(i)}=0,\text{ \ \ }i=0,1,3.
\label{deformed back field}
\end{equation}%
The advantage of this choice lies in the fact that the matter and ghost sectors decouple.

Now, from the equations of motion \eqref{ghost eom} and \eqref{k-def eom}, we obtain the operators
governing the fluctuations of the bosonic and fermionic sectors. For the
bosonic sector we get
\begin{equation}
\mathcal{D}_{B}(x)=\left( 
\begin{array}{cccccc}
\partial _{-}-\hat{n}_{-} & -\lambda^{-2}n_{+} & 0 & -\theta _{+} & \lambda^{-2} \theta _{-} & \theta _{+}
\\ 
-\lambda^{-2}\hat{n}_{-} & \partial _{+}-n_{+} & -\theta _{-} & 0 & \theta _{-} & \lambda ^{-2} \theta _{+}
\\ 
-\theta _{-} & \theta _{+} & -\partial _{-} & \partial _{+} & -s \hat{n}_{-} & s n_{+} \\ 
0 & 0 & \partial _{-} & \partial _{+} & 0 & 0 \\ 
0 & 0 & 0 & -n_{+} & \partial_{-}-\hat{n}_{-} & n_{+} \\ 
0 & 0 & -\hat{n}_{-} & 0 & \hat{n}_{-} & \partial _{+}-n_{+}%
\end{array}%
\right)  \text{ \ \ acting on \ \ }\left( 
\begin{array}{c}
\delta I_{+}^{(2)} \\ 
\delta I_{-}^{(2)} \\ 
\delta I_{+}^{(0)} \\ 
\delta I_{-}^{(0)} \\
\delta N^{(0)}_{+} \\
\delta \hat{N}^{(0)}_{-}
\end{array}%
\right) , \label{det bos x}
\end{equation}%
where the fourth line from top to bottom corresponds to the gauge fixing condition of the remnant gauge symmetry. For the
fermionic sector, we obtain
\begin{equation}
\mathcal{D}_{F}(x)=\left( 
\begin{array}{cccc}
\partial _{-}-\hat{n}_{-} & -\lambda^{-2} n _{+} &\theta _{-} & -\theta _{+} \\ 
-\lambda^{-2} \hat{n} _{-} & \partial _{+}-n_{+} & 0 & 0 \\ 
0 & 0 & \partial _{-}-\hat{n}_{-} & -\lambda^{-2} n _{+} \\ 
-\theta _{-} & \theta _{+} & -\lambda^{-2} \hat{n} _{-} & \partial _{+}-n_{+}%
\end{array}%
\right) \text{ \ \ acting on \ \ }\left( 
\begin{array}{c}
\delta I_{+}^{(1)} \\ 
\delta I_{-}^{(1)} \\ 
\delta I_{+}^{(3)} \\ 
\delta I_{-}^{(3)}%
\end{array}%
\right) . \label{det fer x}
\end{equation}

The 1-loop contribution to the effective Lagrangian, in Euclidean signature, is
\begin{equation}
\mathcal{L}_{E}^{(1)}=\frac{1}{2}\dint\nolimits_{|p|<\mu }\frac{d^{2}p}{%
(2\pi )^{2}}tr[\log \mathcal{D}_{B}(p)-\log \mathcal{D}_{F}(p)],
\end{equation}
where
\begin{equation}
\mathcal{D}_{B}(p)=\left( 
\begin{array}{cccccc}
p_{-}-\hat{n} _{-} & -\lambda^{-2}n_{+} & 0 & -\theta _{+} & \lambda^{-2} \theta _{-} & \theta _{+}
\\ 
-\lambda^{-2}\hat{n} _{-} & p_{+}-n_{+} & -\theta _{-} & 0 & \theta _{-} & \lambda ^{-2} \theta _{+}
\\ 
-\theta _{-} & \theta _{+} & -p_{-} & p_{+} & -s \hat{n} _{-} & s n_{+} \\ 
0 & 0 & p_{-} & p_{+} & 0 & 0 \\ 
0 & 0 & 0 & -n_{+} & p_{-}-\hat{n} _{-} & n_{+} \\ 
0 & 0 & -\hat{n} _{-} & 0 & \hat{n} _{-} & p_{+}-n_{+}%
\end{array}%
\right)
\end{equation}%
and
\begin{equation}
\mathcal{D}_{F}(p)=\left( 
\begin{array}{cccc}
p_{-}-\hat{n} _{-} & -\lambda^{-2} n _{+} &\theta _{-} & -\theta _{+} \\ 
-\lambda^{-2} \hat{n} _{-} & p_{+}-n_{+} & 0 & 0 \\ 
0 & 0 & p _{-}-\hat{n} _{-} & -\lambda^{-2} n _{+} \\ 
-\theta _{-} & \theta _{+} & -\lambda^{-2} \hat{n} _{-} & p_{+}-n_{+}%
\end{array}%
\right).
\end{equation}
The contributions associated to the logarithmic divergences (denoted by $\overset{\cdot }{=}$) are
\begin{equation}
\begin{aligned}
\frac{1}{2}\dint\nolimits_{|p|<\mu }\frac{d^{2}p}{(2\pi )^{2}}tr[\log 
\mathcal{D}_{B}(p)]&\overset{\cdot }{=}-\frac{1}{8\pi }\ln \mu \lbrack
Tr_{adj}^{(0)}+Tr_{adj}^{(2)}]\left( \theta _{+}\theta _{-}+2\lambda
^{-4}n_{+}\hat{n} _{-}\right) , \\
\frac{1}{2}\dint\nolimits_{|p|<\mu }\frac{d^{2}p}{(2\pi )^{2}}tr[\log 
\mathcal{D}_{F}(p)]&\overset{\cdot }{=}-\frac{1}{8\pi }\ln \mu \lbrack
Tr_{adj}^{(1)}+Tr_{adj}^{(3)}]\left( \theta _{+}\theta _{-}+2\lambda
^{-4}n_{+}\hat{n} _{-}\right) .
\end{aligned}
\end{equation}
Then, $\mathcal{L}^{(1)}_{E}\approx c_{2}(\mathfrak{f})=0$, because of the dual Coxeter number $c_{2}(\mathfrak{f})$ of $\mathfrak{f}=\mathfrak{psu}(2,2|4)$ vanishes. As a consequence, the deformation preserves the 1-loop conformal invariance of its parent theory. The same fate is found in the Green-Schwarz formalism in the $AdS_{5}\times S^{5}$ lambda background \cite{Tim-beta} and the hybrid formalism in the $AdS_{2}\times S^{2}$ lambda background \cite{exploring}.

\subsection{Relating the lambda deformed PS/GS background fields}

To compute the target space supergeometry, the gauge fields $A_{\pm}$ must be integrated out completely. Then, after using the gauge fields equations of motion we obtain the following effective action
\begin{equation}
S_{\text{eff}}=S_{\text{matter}}+S_{\text{ghost}}, \label{eff act}
\end{equation}%
where%
\begin{equation}
\begin{aligned}
S_{\text{matter}} &=-\frac{k}{2\pi }\int\nolimits_{\Sigma }d^{2}\sigma \left\langle \mathcal{F%
}^{-1}\partial _{+}\mathcal{F}\left\{ 1+2\mathcal{O}^{-1}D\right\} \mathcal{F%
}^{-1}\partial _{-}\mathcal{F}\right\rangle +S_{WZ}, \\
S_{\text{ghost}} =&-\frac{k}{\pi}s\int\nolimits_{\Sigma }d^{2}\sigma
\big\langle \hat{N}_{-}^{(0)}\mathcal{O}^{-T}\mathcal{F}^{-1}\partial _{+}%
\mathcal{F-}N_{+}^{(0)}\mathcal{O}^{-1}\partial _{-}\mathcal{FF}%
^{-1}-sN_{+}^{(0)}\mathcal{O}^{-1}\hat{N}_{-}^{(0)}\big\rangle 
\\
&-\frac{k}{\pi}s\int\nolimits_{\Sigma }d^{2}\sigma \big\langle
w_{+}^{(3)}\partial _{-}l^{(1)}+\hat{w}_{-}^{(1)}\partial _{+}\hat{l}%
^{(3)}-N_{+}^{(0)}\hat{N}_{-}^{(0)}\big\rangle \label{eff act 2}
\end{aligned}
\end{equation}%
and $\mathcal{O}$ as defined in \eqref{O}. This action is, as expected, invariant under the (on-shell) BRST symmetry variations \eqref{on-shell BRST}.

From the expressions of the projectors in the PS and the GS lambda models, i.e.
\begin{equation}
\begin{aligned}
\Omega  &=P^{(0)}+\lambda ^{-3}P^{(1)}+\lambda ^{-2}P^{(2)}+\lambda
^{-1}P^{(3)}, \\
\Omega _{gs} &=P^{(0)}+\lambda P^{(1)}+\lambda ^{-2}P^{(2)}+\lambda
^{-1}P^{(3)},
\end{aligned}
\end{equation}%
we get an important relation between both formalisms
\begin{equation}
\mathcal{O=O}_{gs}+s\lambda P^{(1)},
\end{equation}%
where%
\begin{equation}
\mathcal{O}_{gs}=\Omega _{gs}-D.
\end{equation}

The strategy for obtaining a clear and direct relation between the target space geometry of the PS and the GS lambda models start by introducing the following GS super-vielbeins defined by 
\begin{equation}
E_{\pm }=\mathcal{O}_{gs}^{-T}\mathcal{F}^{-1}\partial _{\pm }\mathcal{F},%
\text{ \ \ }\hat{E}_{\pm }=\mathcal{O}_{gs}^{-1}\partial _{\pm }\mathcal{%
FF}^{-1}. \label{vielbeins}
\end{equation}%
This strategy was used successfully in \cite{eta-PS} for understanding the relation between the Yang-Baxter deformations of the PS and the GS formulations of the $AdS_{5}\times S^{5}$ superstring. Our purpose here is to apply the same approach to the case at hand.  

Consider now the quantity%
\begin{equation}
X=\mathcal{O}_{gs}\left( 1+2\mathcal{O}^{-1}D\right) \mathcal{O}_{gs}^{T}.
\end{equation}%
The metric and the antisymmetric fields are extracted, respectively, from the symmetric and the anti-symmetric parts of $X$, namely
\begin{equation}
\begin{aligned}
G' &=\mathcal{O}_{gs}\left( 1+\mathcal{O}^{-1}D+D^{T}\mathcal{O}^{-T}\right) 
\mathcal{O}_{gs}^{T} \\
&=(\Omega _{gs}\Omega _{gs}^{T}-1)-s\lambda \big\{ P^{(1)}%
\mathcal{O}^{-1}D\mathcal{O}_{gs}^{T}+\mathcal{O}_{gs}D^{T}\mathcal{O}%
^{-T}P^{(3)}\big\}  \\
&=s\big[ P^{(2)}-\lambda \big\{ P^{(1)}\mathcal{O}^{-1}D%
\mathcal{O}_{gs}^{T}+\mathcal{O}_{gs}D^{T}\mathcal{O}^{-T}P^{(3)}\big\} %
\big] 
\end{aligned}
\end{equation}
and%
\begin{equation}
\begin{aligned}
B' &=\mathcal{O}_{gs}\left( \mathcal{O}^{-1}D-D^{T}\mathcal{O}^{-T}\right) 
\mathcal{O}_{gs}^{T} \\
&=\left( D\Omega _{gs}^{T}-\Omega _{gs}D^{T}\right) -s\lambda
\big\{ P^{(1)}\mathcal{O}^{-1}D\mathcal{O}_{gs}^{T}-\mathcal{O}_{gs}D^{T}%
\mathcal{O}^{-T}P^{(3)}\big\} .
\end{aligned}
\end{equation}
Then, we have that
\begin{equation}
\left\langle \mathcal{F}^{-1}\partial _{+}\mathcal{F}\left\{ 1+2\mathcal{O}%
^{-1}D\right\} \mathcal{F}^{-1}\partial _{-}\mathcal{F}\right\rangle
=\left\langle E_{+}(G'+B')E_{-}\right\rangle ,
\end{equation}%
where%
\begin{equation}
\begin{aligned}
\left\langle E_{+}G'E_{-}\right\rangle  &=s\langle
E_{+}^{(2)}E_{-}^{(2)}- \{ E_{+}^{(3)}P\hat{E}_{-}^{(1)} +E_{-}^{(3)}P\hat{E}_{+}^{(1)} \} \rangle,  \\
\left\langle E_{+}B'E_{-}\right\rangle  &=\left\langle E_{+}\left( D\Omega
_{gs}^{T}-\Omega _{gs}D^{T}\right) E_{-}\right\rangle -s
\langle E_{+}^{(3)}P \hat{E}%
_{-}^{(1)} -E_{-}^{(3)}P\hat{E}%
_{+}^{(1)}\rangle 
\end{aligned}
\end{equation}
and where we have defined\footnote{Notice that $P:\mathfrak{f}^{(1)}\rightarrow \mathfrak{f}^{(1)}$.}
\begin{equation}
P=\lambda \mathcal{O}_{gs} \mathcal{O}^{-1}P^{(1)}.
\end{equation}

Form these results, the matter contribution to the effective action take the form
\begin{equation}
S_{\text{matter}}=-\frac{k}{2\pi }s\int\nolimits_{\Sigma
}d^{2}\sigma \big\langle E_{+}^{(2)}E_{-}^{(2)}+E_{+}BE_{-}-2E_{+}^{(3)}P 
\hat{E}_{-}^{(1)}\big\rangle ,
\end{equation}%
where%
\begin{equation}
B=s^{-1}B_{0}+(D\Omega _{gs}^{T}-\Omega _{gs}D^{T}).
\end{equation}%
Above, we have written the WZ term, locally, in the form%
\begin{equation}
S_{WZ}=-\frac{k}{2\pi }\int\nolimits_{\Sigma }d^{2}\sigma \left\langle
E_{+}B_{0}E_{-}\right\rangle.
\end{equation}

Before focusing on the ghost contribution, consider first the following two identities
\begin{equation}
\begin{aligned}
P^{(0)}\mathcal{O}^{-T}\mathcal{O}_{gs}^{T} &=P^{(0)}-s\lambda P^{(0)}\mathcal{O}^{-T}P^{(3)} \\
&=P^{(0)}+\hat{C}\hat{P}, \\
P^{(0)}\mathcal{O}^{-1}\mathcal{O}_{gs} &=P^{(0)}-s\lambda
P^{(0)}\mathcal{O}^{-1}P^{(1)} \\
&=P^{(0)}-CP,
\end{aligned}
\end{equation}
where\footnote{Notice that $\hat{P}:\mathfrak{f}^{(3)}\rightarrow \mathfrak{f}^{(3)}$ and $C,\hat{C}\neq C^{T}:\mathfrak{f}\rightarrow \mathfrak{f}^{(0)}$.}%
\begin{equation}
\hat{P}=P^{T}=\lambda \mathcal{O}_{gs}^{T}\mathcal{O}^{-T}P^{(3)},\text{ \ \ }%
C=sP^{(0)}\mathcal{O}_{gs}^{-1},\text{ \ \ }\hat{C}%
=-sP^{(0)}\mathcal{O}_{gs}^{-T}.
\end{equation}
It follows from these expressions that
\begin{equation}
\begin{aligned}
\big\langle N_{-}^{(0)}\mathcal{O}^{-T}\mathcal{F}^{-1}\partial _{+}%
\mathcal{F}\big\rangle  &=\big\langle N_{-}^{(0)}\big(E_{+}^{(0)}+%
\hat{C}\hat{P}E_{+}^{(3)}\big) \big\rangle , \\
\big\langle N_{+}^{(0)}\mathcal{O}^{-1}\partial _{-}\mathcal{FF}%
^{-1}\big\rangle  &=\big\langle N_{+}^{(0)}\big(\hat{E}%
_{-}^{(0)}-CP\hat{E}_{-}^{(1)}\big) \big\rangle .
\end{aligned}
\end{equation}

Using these results, we have the ghost contribution to the effective action
\begin{equation}
S_{\text{ghost}}=-\frac{k}{\pi }s\int\nolimits_{\Sigma
}d^{2}\sigma
\big\langle N_{+}^{(0)}\big( \hat{\Theta }_{-}+CP\hat{E%
}_{-}^{(1)}\big)+ \hat{N}_{-}^{(0)}\big( \Theta_{+} +\hat{C}\hat{P}%
E_{+}^{(3)}\big) +N_{+}^{(0)}S\hat{N}_{-}^{(0)}\big\rangle +S_{lw},
\end{equation}%
where $S_{lw}$ is given by the usual ghost term
\begin{equation}
S_{lw}=-\frac{k}{\pi}s\int\nolimits_{\Sigma }d^{2}\sigma \big\langle
w_{+}^{(3)}\partial _{-}l^{(1)}+\hat{w}_{-}^{(1)}\partial _{+}\hat{l}%
^{(3)}\big\rangle
\end{equation}
and
\begin{equation}
\begin{aligned}
S &=-P^{(0)}-sP^{(0)}\mathcal{O}^{-1}P^{(0)},\text{ \ \ } \Theta_{+}  &=E_{+}^{(0)},\text{ \ \ }\hat{\Theta }_{-}=-\hat{E}_{-}^{(0)}.
\end{aligned}
\end{equation}

Remarkably, the effective action of the lambda deformed PS superstring \eqref{eff act} takes the standard form of the Berkovits-Howe (BH) action functional \cite{Ber-How}
\begin{equation}
S_{\text{BH}}=-T\int\nolimits_{\Sigma }d^{2}\sigma L_{\text{BH}},\text{ \ \ }%
T=\frac{k}{\pi }(\lambda ^{-4}-1), 
\end{equation}%
where
\begin{equation}
\begin{aligned}
L_{\text{BH}} =\, &\big\langle\frac{1}{2}E_{+}^{(2)}E_{-}^{(2)}+\frac{1}{2}%
E_{+}BE_{-}-E_{+}^{(3)}P\hat{E}_{-}^{(1)}+N_{+}^{(0)}\big( \hat{%
\Theta }_{-}+CP\hat{E}_{-}^{(1)}\big)  \\
&+\hat{N}_{-}^{(0)}\big( \Theta _{+}+\hat{C}\hat{P}%
E_{+}^{(3)}\big) +N_{+}^{(0)}S\hat{N}_{-}^{(0)}\big\rangle+L_{lw}. \label{BH Lagrangian}
\end{aligned}
\end{equation}
The BH action is the most general action functional which possesses BRST symmetry, classical world-sheet conformal symmetry and zero ghost number. As argued in \cite{Ber-How}, the ten-dimensional supergravity constraints are a consequence of the BRST symmetry\footnote{In constrast, for the Green-Schwarz formulation the supergravity constraints are a consequence of kappa symmetry, a condition also satisfied by the lambda deformation of the Green-Schwarz $AdS_{5}\times S^{5}$ superstring \cite{lambda-fer}.} of the action \eqref{BH Lagrangian} and because of the action \eqref{eff act} fulfils this condition, it describes a supergravity background.  

Once the action is in this canonical form, the background fields are easily identified and in our case they are encoded in the objects of the following list:
\begin{equation}
\begin{aligned}
\text{Background metric} &: P^{(2)}, \\ 
\text{Superspace two-form} &: B, \\ 
\text{Ramond-Ramond bispinor} &: P, \\ 
\text{Gravitini and dilatini} &: C,\hat{C}, \\ 
\text{Left/right moving spin connection} &: \Theta_{+},\hat{\Theta }_{-} \\
\text{Riemann curvature} &: S.%
\end{aligned}
\end{equation}
The vielbeins, the metric, the $B$-field, the RR-bispinor $P$ (actually its inverse in index notation) and the spin connections $\Theta_{+}$, $\hat{\Theta }_{-}$ are equal to 
the SUGRA background fields of the $AdS_{5}\times S^{5}$ Green-Schwarz lambda model found in \cite{Borsato}. The $C$, $\hat{C}$ and $S$ are auxiliary fields, i.e. can be defined in terms of the vielbeins \cite{Ber-How,Mazzucato}. 

Thus, we conclude that both models (the lambda deformations of the GS and PS superstring) describe the same type IIB supergravity background and this means that our action \eqref{eff act} corresponds to the pure spinor formulation of the $AdS_{5}\times S^{5}$ superstring in the lambda background. In order to accomplish the correct equations of motion for type IIB supergravity, the dilaton must be the same as that obtained for the lambda model of the GS superstring \cite{Borsato}. However, the contribution from integrating out the gauge fields $A_{\pm}$, gives rise to the following dilaton field 
\begin{equation}
e^{-2\phi }=sdet\,\mathcal{O}_{},
\end{equation} 
which is different from the GS dilaton as both are related via 
\begin{equation}
 e^{-2\phi }=sdet(\mathcal{O}_{gs}^{-1}\mathcal{O})e^{-2\phi_{gs} }.
\end{equation}
Then, for consistency with the supergravity equations of motion, the extra term
\begin{equation}
\mathcal{J}=sdet(\mathcal{O}_{gs}^{-1}\mathcal{O})^{-1/2},
\end{equation}
should be interpreted as a Jacobian determinant arising from a change of group field variables in the path integral Haar measure\footnote{This is reasonable, as the group field $\mathcal{F}$ intrinsically depend on the deformation parameter $\lambda$ and this dependence might differ on both formulations. } $\mathcal{DF} \rightarrow \mathcal{DF}^{\prime}$. 

Finally, we write the BRST currents \eqref{chiral BRST} in the standard curved space form \cite{Mazzucato}
\begin{equation}
j_{+} =\big\langle l^{(1)}d^{(3)}\big\rangle,\text{ \ \ } j_{-} =\big\langle \hat{l}^{(3)}\hat{d}^{(1)}\big\rangle,
\end{equation}
in terms of the world-sheet auxiliary fields
\begin{equation}
d^{(3)} =\hat{P}\big( E_{+}^{(3)}-C^{T}N_{+}^{(0)}\big),\text{ \ \ }
\hat{d}^{(1)} =-P\big( \hat{E}_{-}^{(1)}-\hat{C}^{T}\hat{N}_{-}^{(0)}\big), \label{auxiliary eom}
\end{equation}
which is the correct form for the BRST chiral currents in the BH approach \cite{Ber-How}.
 
\section{Conclusions}

We have shown how to lambda deform the pure spinor formalism of the superstring in the $AdS_{5}\times S^{5}$ background in a consistent way. The deformation preserves the BRST symmetry, the classical integrability, the local symmetries and the 1-loop conformal symmetry of its parent theory. Furthermore, the target space supergeometry is exactly the same as that of the lambda model of the Green-Schwarz formulation of the superstring and satisfy the same set of supergravity equations of motion. This result complements an analogous equivalence, found recently, between the Yang-Baxter deformations of the PS and GS formulations of the superstring in the $AdS_{5}\times S^{5}$ background.

Concerning the quantum conformal symmetry of the deformed theory, some comments are in order. For the $AdS$ background, it has been argued that the effective action of the un-deformed theory is conformal invariant to all-loop \cite{Berkovits}. This is proven by using cohomological arguments and the fact that the cohomology of the BRST charge at ghost number 1 is empty. In this sense, it would be interesting to understand whether this result can be generalized for the $\lambda$-deformed BRST charge. In \cite{Bedoya-Chandia} it was shown that the one-loop beta function for the type II pure spinor superstring in a curved background vanish as a consequence of the BRST invariance of the Berkovits-Howe action. The conformal invariance at all-loop has never been proven and, in the particular case of the $\lambda$-deformation, because of the fact that we still do not have a diagrammatic proof of conformal invariance at all-loop level this is a challenging problem clearly out of the scope of the present contribution. For these reasons, we shall content ourselves with the one-loop proof of conformal symmetry provided above and leave the general problem for a future work.

It would be interesting to study as well the relation between the PS formalism of the $AdS_{5}\times S^{5}$ superstring and the Chern-Simons theories considered in \cite{lambdaCS,lambdaCS2}. In case of a consistent relation, the non-ultralocality term present in the algebra \eqref{Maillet-lambda} of the PS formalism could be eliminated (for any value of the deformation parameter $\lambda$) with the added advantages of not having to deal neither with the kappa symmetry nor the light-cone gauge. This will be considered in a companion paper.

\section*{Acknowledgements}

The work of DMS is supported by the S\~ao Paulo Research Foundation (FAPESP) under the research grant 2017/25361-7. The authors thank the referee for the very useful comments and suggestions.


\end{document}